# Bolt Detection Signal Analysis Method Based on ICEEMD

**Chunhui Guo[1,2], Zhan Zhang[1*], Xin Xie[3], Zhengyu Yang[3]**

[1.] College of Water Conservancy and Hydropower Engineering, Hohai University, Nanjing, China

[2.] Key Laboratory of Hydraulic and Waterway Engineering of the Ministry of Education, Chongqing Jiaotong University, China

[3.] Department of Electrical and Computer Engineering, Northeastern University, Boston, USA

[*]Corresponding author

E-mail:*337221231@qq.com,zhanzhang_hhu@qq.com,xie.x@husky.neu.edu,yang.zhe@husky.neu.edu*

**Abstract:**

The construction quality of the bolt is directly related to the safety of the project, and as such, it must be tested. In this paper, the improved complete ensemble empirical mode decomposition (ICEEMD) method is introduced to the bolt detection signal analysis. The ICEEMD is used in order to decompose the anchor detection signal according to the approximate entropy of each intrinsic mode function (IMF). The noise of the IMFs is eliminated by the wavelet soft threshold de-noising technique. Based on the approximate entropy, and the wavelet de-noising principle, the ICEEMD-De anchor signal analysis method is proposed. From the analysis of the vibration analog signal, as well as the bolt detection signal, the result shows that the ICEEMD-De method is capable of correctly separating the different IMFs under noisy conditions, and also that the IMF can effectively identify the reflection signal of the end of the bolt.

**Keywords**：bolt anchorage, approximate entropy, intrinsic mode function, wavelet de-noise, CEEMD

## 1. Introduction

The bolt anchoring system is subject to the geological conditions and the construction technology effect. If there are any hidden dangers that haven't been detected, then it will cause engineering accidents and serious economic losses. Therefore, the construction quality of the bolt anchorage must be checked, so as to ensure the safety of the project. During the early stage, the detection of the anchor's anchoring quality is mainly based on the drawing test [1-3]. However, this method will cause damage to the anchoring system. The method is also not suitable for large-scale detection and can't be fully reflected [4-6].

The detection method for the quality of the anchor will be gradually replaced by the use of non-destructive testing methods, such as the acoustic wave method [7-10]. This method is established based on the mathematical model of the one-dimensional elastic rod [11-13]. The assumption is that the longitudinal wave wavelength that is generated by the exciting force is much larger than that of the bolt radius, so the transverse displacement of the system can be neglected. By solving the longitudinal one-dimensional wave equation, the dynamic response of the bolt system is obtained.

The low-end reflection signal of the bolt can be easily disturbed during the process of bolt detection; it is difficult to directly obtain the reflected wave arrival time. In order to obtain the effective signal, many data processing methods, such as the short-time Fourier transform, the Gabor transform, the Wigner-Ville transform, the wavelet transform and so on, are proposed. Wavelet transform is the most used signal analysis method among them[14-18]. However, the effect of the wavelet transform is often limited by the wavelet base, as well as the number of decomposed layers.

The empirical mode decomposition (EMD) can adaptively select the substrate according to the characteristics of the signal for the multi-resolution analysis of the signal, which will overcome the wavelet base selection problem [19-21]. The decomposition is based on the local timescale of the data. There have been many applications about EMD processing detection signal [22-25]. However, the EMD encounters some modal aliasing problems during the processing procedure [26-28]. The Ensemble empirical mode decomposition (EEMD) overcomes the modal aliasing problem that's inherent of the EMD, but due to the addition of different white noise, the decomposition may produce a false mode, which can also cause errors. The reconstructed signal still includes residual noise, and different realizations of signal and noise may produce different modes [29-31]. Complete EEMD (CEEMD) has been successfully applied to seismic signal analysis. Han et al. used CEEMD to obtain an exact reconstruction of the original signal and a better spectral separation of the modes with synthetic and real seismic data [32]. However, the CEEMD cannot be proven, and the final averaging problem remains unsolved since different noisy copies of the signal can produce a different number of modes [33]. In recent years, the improved complete EEMD (ICEEMD) has been proposed by adding adaptive white noise to the signal and by redefining the calculations of the local mean for each model [33-37]. The result shows that the method is superior to the traditional method. Although EMD is more adaptive and more efficient [38, 39], the EEMD outperforms EMD in causing less mode mixing [40, 41]. The CEEMD outperforms EMD in causing less mode mixing and EEMD in better reconstruction performance [42, 43]. The ICEEMD, as illustrated in the paper, outperforms CEEMD in being more physical meaningful and less number of components [44].

Based on the mentioned research, the ICEEMD method is introduced into the bolt detection signal analysis in this paper. However, the actual signal of bolt detection is under noise interference. The processing signal under noise is critical problem with ICEEMD for bolt detection. By combining the approximate entropy and the wavelet de-noising principle, the ICEEMD-De was established based on the ICEEMD. Then, the ICEEMD-de was used to process the simulation vibration signal and the actual bolt detection signal.

## 2. Theory and Methodology

Based on the ICEEMD anchor detection signal analysis method, the ICEEMD-De integrates ICEEMD, the approximate entropy and wavelet de-noising. The three methods are introduced in the section.

2.1. ICEEMD Principle

The ICEEMD method is able to effectively prevent the occurrence of false IMF by adding the adaptive white noise to the signal and by redefining the local mean of each modal. Assuming the anchor detection signal $s$, then the decomposition process of the ICEEMD is as follows:

1）The signal $s$ is added to the M group Gaussian white noise in order to generate a new signal $s^i$. $s^i$ can be expressed as:

$$s^i = s + \beta_k w^i \quad (1)$$

where $w^i$ ($i=1,2,…,M$) is one group of Gaussian white noise, $\beta_k = \varepsilon_0 std(r_k)$, $r_k$ is the $k$-th residue, $\varepsilon_0$ takes 0.2.

2）The $k$-th mode can be obtained by EMD. We can obtain the mean of the $k$-th mode and have:

$$\langle E_k(s^i) \rangle = s^i - \langle M(s^i) \rangle \quad (2)$$

where $\langle \ \rangle$ is the operator of mean.

3) $s^i$ is decomposed by using EMD. We obtain 1-th residue $r_1$ and 1-st IMF $d_1$. We have:

$$r_1 = \langle M(s^i) \rangle \quad (3)$$

$$d_1 = s - r_1 \quad (4)$$

4) We take 2-nd residue, $r_2$, as the local mean of $r_1 + \beta_1 E_2(w^j)$. The 2-nd IMF $d_2$ is:

$$r_2 = \langle M(r_1 + \beta_1 E_2(w^j)) \rangle \quad (5)$$

$$d_2 = r_1 - r_2 \quad (6)$$

5) For any $r_k$, and k-th IMF $d_k$, the expression is as follows:

$$r_k = \langle M(r_{k-1} + \beta_{k-1} E_k(w^j)) \rangle \quad (7)$$

$$d_k = r_{k-1} - r_k \quad (8)$$

Go to step 3), we obtain all of the IMF.

2.2. Approximate Entropy

All of the IMF approximate entropy can be expressed as $\{A\} = \{A_1, A_2, \cdots, A_k\}$. Then the calculation procedure $A_k$ is as follows:

1) Take k-th IMF as the time series of n points and define it as:

$$\{Z\} = \{z_1, z_2, \cdots, z_n\} \quad (9)$$

2) Compute the binary distance matrix B of the time series:

$$B = \begin{bmatrix} b_{11} & b_{11} & \cdots & b_{1n} \\ b_{21} & b_{22} & \cdots & b_{2n} \\ \vdots & \vdots & & \vdots \\ b_{n1} & b_{n2} & \cdots & b_{nn} \end{bmatrix} \quad (10)$$

where $b_{ij} = \begin{cases} 1, |z_i - z_j| < a \\ 0, 1, |z_i - z_j| \geq a \end{cases}$, $a$ is threshold ($a$=0.1~0.2) [45].

3) Compute the ratio of n+m+1 and n-m+2 to number for B matrix element less than $a$ $C_i^2$ and $C_i^3$ is as following:

$$C_i^2 = \sum b_{ij} \cap b_{(i+1)(j+1)}, j = 1, 2, \cdots, n-1 \quad (11)$$

$$C_i^3 = \sum b_{ij} \cap b_{(i+1)(j+1)} \cap b_{(i+2)(j+2)}, j=1,2,\cdots,n-2 \qquad (12)$$

4) Compute the $C_i$ of the nature logarithm, and get the average of the $C_i$ of the nature logarithm. Then the approximate entropy of the k-th IMF $A_k$ is as follows:

$$\Phi_1 = \frac{1}{n-m+1} \sum_{i=1}^{n-m+1} \ln C_i^2 \qquad (13)$$

$$\Phi_2 = \frac{1}{n-m+2} \sum_{i=1}^{n-m+1} \ln C_i^3 \qquad (14)$$

$$A_k = \Phi_1 - \Phi_2 \qquad (15)$$

5) The wavelet de-noise is performed with the approximate entropy of the IMF that is greater than the threshold.

2.3. Wavelet De-noise

The wavelet de-noise is achieved based on a critical threshold. The main steps of its de-noise principle are as follows:

1) Select the appropriate wavelet base and the number of decomposition layers. We take the wavelet transform with the noise signal $s$ and obtain its wavelet coefficients $w^j$:

$$v^{j-1} = D_0 H v^j \qquad (16)$$

$$w^{j-1} = D_0 G v^j \qquad (17)$$

Where $H$ is the low-pass filter, $G$ is the high-pass filter, $v$ is for the scale factor, and $w$ is the wavelet coefficient.

2) Select the appropriate threshold function to process the wavelet coefficients $w^j$ and get the estimation wavelet coefficients $\hat{w}^j$:

$$\begin{cases} \hat{w}^j = \text{sgn}(w^j)(|w^j| - \lambda), & |w^j| \geq \lambda \\ \hat{w}^j = 0, & |w^j| \leq \lambda \end{cases} \qquad (18)$$

where $\lambda$ is threshold, $\lambda = \sigma\sqrt{2\log N}$, $\sigma$ is the mean square error of the signal, $N$ is the sampling point number.

3) use the estimation wavelet coefficients $\hat{w}^j$ and get the reconstruct signal $\hat{v}$

$$\hat{v}^j = \tilde{H} U \hat{w}^{j-1} + \tilde{G} U v^{j-1} \qquad (19)$$

where $\tilde{H}$ is reconstruct low-pass filter, $\tilde{G}$ is reconstruct high-pass filter.

Based on ICEEMD, the approximate entropy and wavelet de-noising theory, we proposed

ICEEMD-De method. The method is divided five steps to implement processing vibrational signal. The ICEEMD-De analysis process is shown in Figure 1. At first step, we sample the anchoring detection signal and take sampling signal for analysis. Then the sampling signal is decomposed with ICEEMD method. Each IMFs of the signal can be obtained. At third step, we solve the approximate entropy of each IMF based on approximate entropy theory. At fourth step, the method takes the approximate entropy as the condition of whether the IMF is de-noised. When approximate entropy of the IMF above threshold, the IMF need take wavelet de-nosed. Finally, by means of the wavelet soft threshold de-noising technique, the noise in the intrinsic mode function (IMF) is eliminated. The original signal components are retained in maximum with ICEEMD-De.

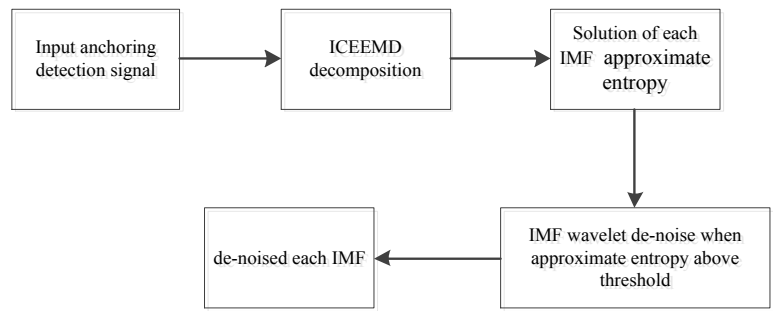

**Fig.1.** Anchoring detection signal analysis flowchart using ICEEMD-De.

## 3. Vibration Simulation Signal Analysis

Focusing on the analysis of the anchor detection signal, one vibration simulation signal is considered, and the signal is decomposed by means of ICEEMD. The ICEEMD method is used in order to analyze the signal under the noise interference condition. The noise signal is directly decomposed by the ICEEMD, ICEEMD decomposition after the wavelet de-noising and ICEEMD-De for studying processing effect.

3.1. Vibration Signal Decompose with ICEEMD

The supposed vibration simulation signal $s$ is composed of the $s_1$ and $s_2$ two sine function (Fig.2.). The expression is as follows:

$$s_1 = \sin(20\pi t) \quad (20)$$

$$s_2 = \begin{cases} 0.4 \times \sin(100\pi t) & \text{when } 0.15 \leq t \leq 0.25 \\ 0 & \text{other} \end{cases} \quad (21)$$

$$s = s_1 + s_2 \quad (22)$$

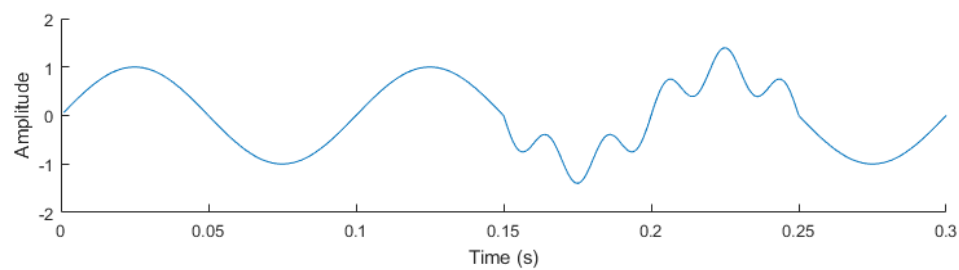

**Fig.2.** Vibration simulation synthetic signal.

The simulation synthetic signal in Fig.2 is decomposed with the ICEEMD. The analysis results in Fig.3 show:

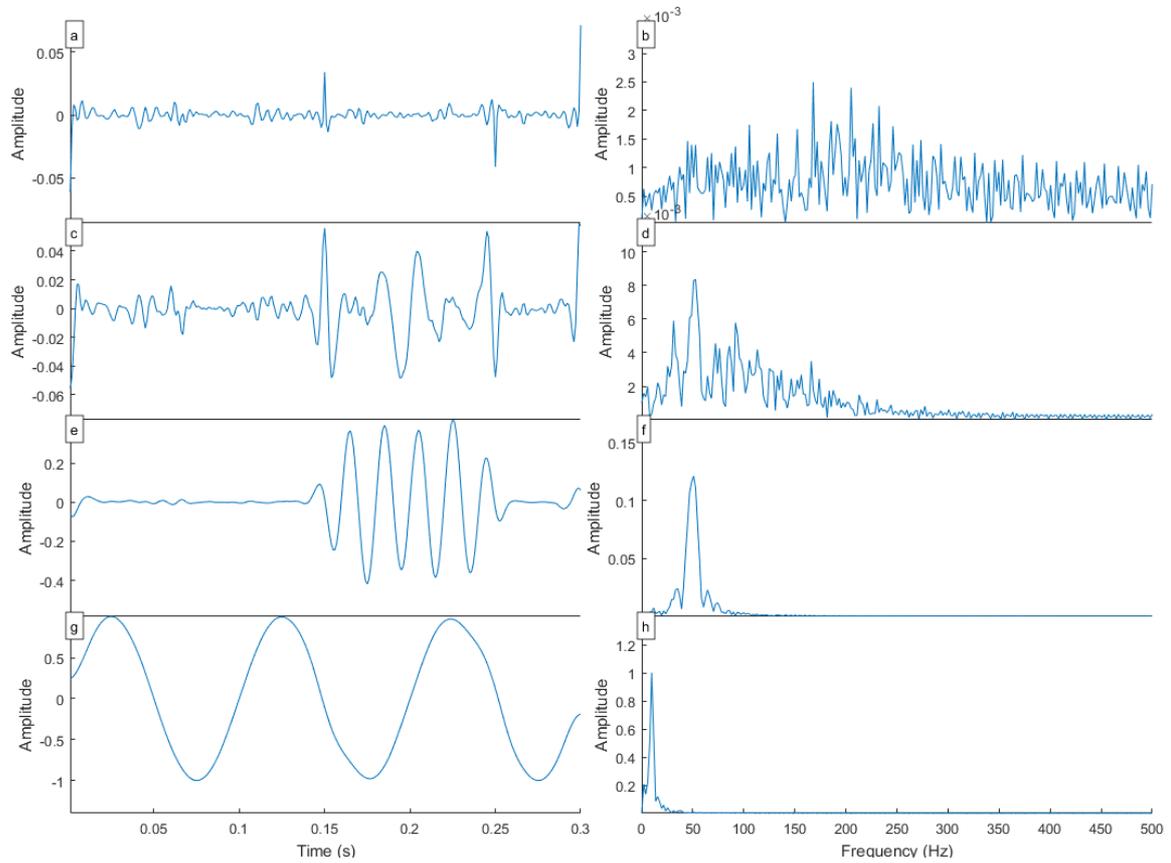

**Fig.3.** IMF and spectrum after the decomposition of the vibration signal with ICEEMD: Fig.3 (a) (c) (e) (g)IMF1 ~ IMF4. Fig3. (b) (d) (f) (h) IMF4 spectrum.

According to Fig. 3, the ICEEMD decomposes the *s* signal into four different IMFs, where the latter two characteristic moduli correspond to $s_2$ and $s_1$, respectively, with the corresponding frequencies of the IMF being clearly seen in regards to the spectrum. Fig 3(b) (d) shows the frequency of both $s_2$ and $s_1$, which are 100Hz and 20Hz, respectively. Fig 3(b) (d) shows that the IMF1 and the IMF2 are mainly high frequency noise signals.

3.2. Vibration Signal Mode Decomposition under Noise Condition with ICEEMD

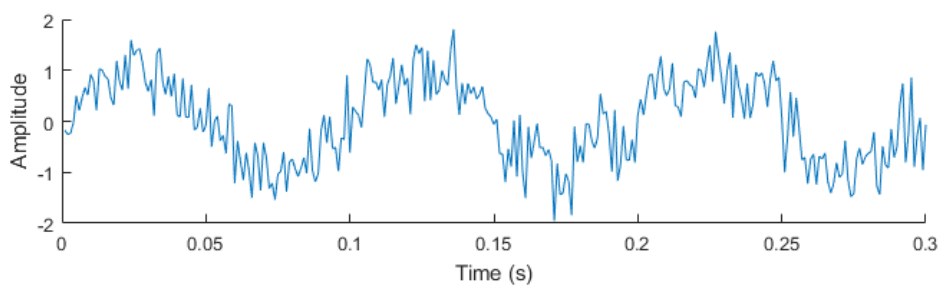

**Fig.4.** Vibration signal under noise condition.

The random signal is added to the source signal in Fig. 2. We take SNR=5dB as example for the analysis. The signal at 5dB is analyzed by means of ICEEMD and ICEEMD-De. The results are shown in Fig.5 and Fig.6.

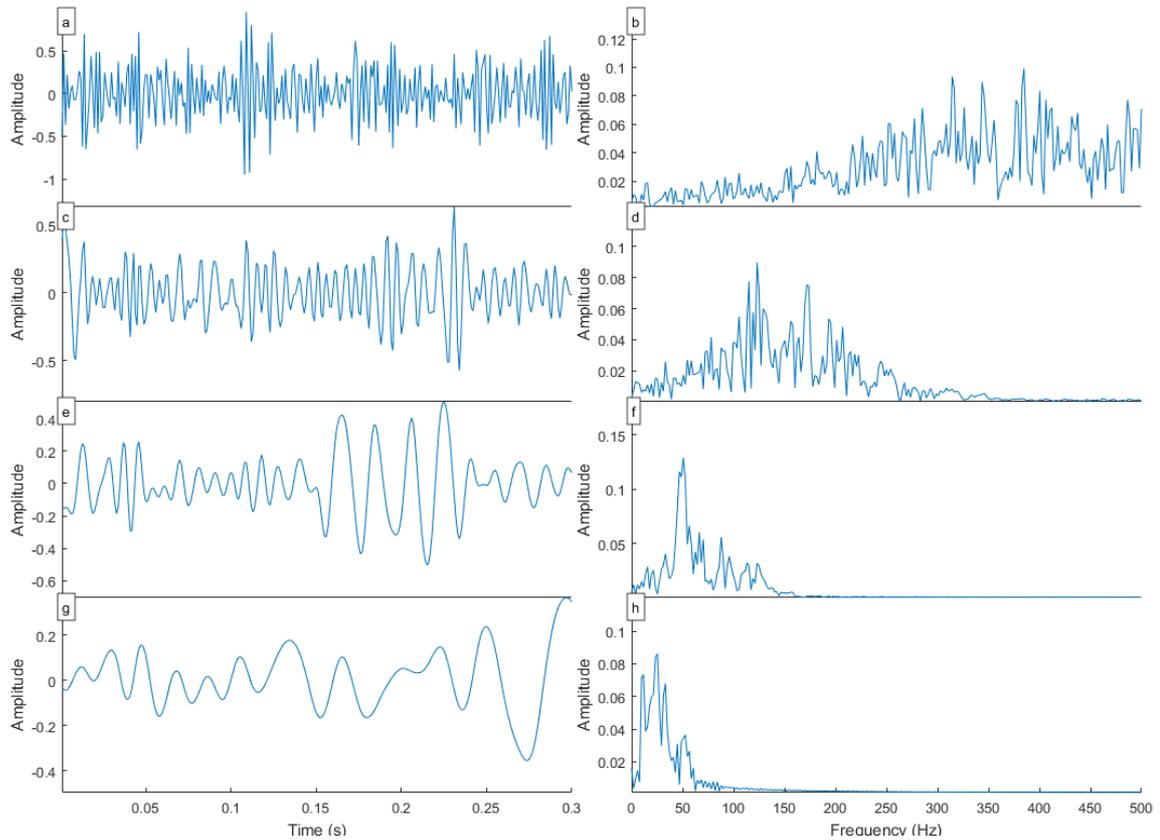

**Fig.5.** The IMF and spectrum which the vibration signal after the decomposition with ICEEMD: Fig.5 (a) (c) (e) (g) is IMF1 ~ IMF4 and Fig.5 (b) (d) (f) (h) is IMF1~IMF4 spectrum.

According to Fig. 5, the IMF1 and IMF2, which were decomposed by ICEEMD, are still dominated by the random noise signal, while the IMF3 and 4 corresponding s1 and s2 are doped with a large number of random interference components.

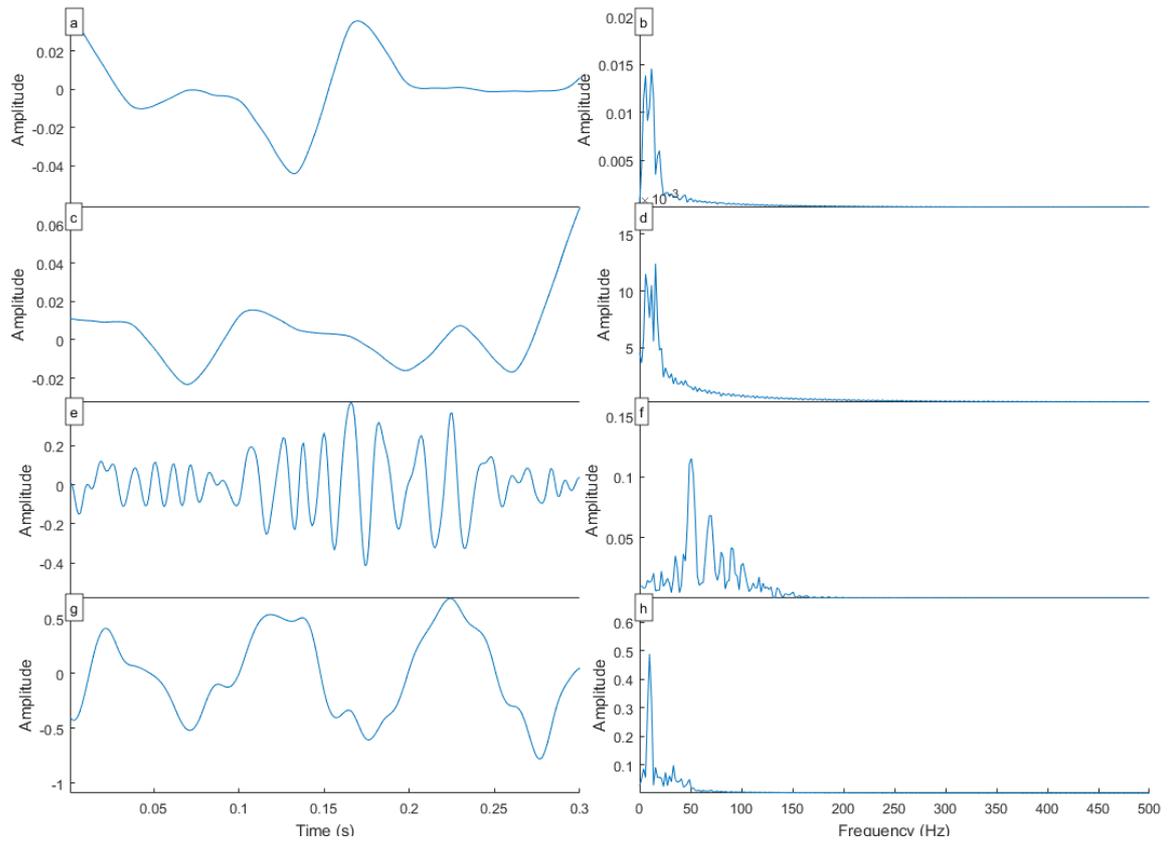

**Fig.6.** The IMF and spectrum which the vibration signal after the decomposition with ICEEMD-De: Fig.5 (a) (c) (e) (g) is IMF1 ~ IMF4 and Fig.5 (b) (d) (f) (h) is IMF1~IMF4 spectrum.

From Fig. 6, it can be seen that a large number of random noise interference in IMF1,2 has removed, and the IMF3,4's corresponding s1, s2, had components of the interference significantly suppressed. The de-noise signal can be clearly distinguished between the s1, s2 frequency of 100Hz and 20Hz.

In order to further study the effect of the proposed method on the de-noising of the vibration signal, the de-noising effect of the wavelet and the ICEEMD-De method on the *s* signal is compared. Fig. 7 shows the error of the vibration signal after the treatment.

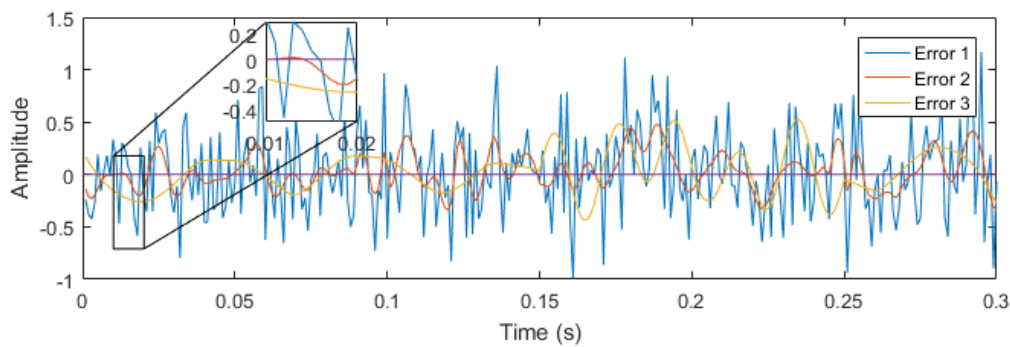

**Fig 7.** De-noising error line with Wavelet and ICEEMD-De.

In Fig. 7, Error1 is the noise error, Error2 is the error after the ICEEMD de-noising and Error3 is the error after the wavelet de-noising. From Fig.7, it can be seen that both methods are able to significantly reduce the noise interference. Based on the general trend of the error line, the proposed method results in the error line's frequency being higher, but with the error size being lower than the wavelet de-noising.

In the signal de-noising analysis, the signal-to-noise ratio (*SNR*) and the root-mean-square deviation （RMSE） are used in order to measure the de-noising effect of the signal, which is defined as follows:

$$SNR = 10\lg\left\{\sum_{i=1}^{N} X_i^2 / \sum_{i=1}^{N}\left(X - \hat{X}_i\right)^2\right\} \quad (23)$$

$$RMSE = \sqrt{\frac{\sum_{n=1}^{N}\left(X_i - \hat{X}_i\right)^2}{N}} \quad (24)$$

According to the formulas(23) (24), two different methods can be calculated in order to de-noise the effect of the index. The index is shown in Table 1.

**Table.1.** ICEEMD de-noising performance index

| Index | Original signal | ICEEMD | Wavelet |
|---|---|---|---|
| SNR (dB) | 5.000 | 13.741 | 11.179 |
| RMSE | 0.411 | 0.151 | 0.201 |

From Table 1, we can see that both methods are able to greatly improve the SNR of the original signal, which is consistent with the results that are shown in Fig 7. The proposed method improves the *SNR* of the original signal by 2.7 times, while the Wavelet method increases the *SNR* of the original signal 2.2 times. The proposed method's RMSE is smaller than that of the Wavelet method. Therefore, the ICEEMD-De de-noising effect is superior to that of the Wavelet de-noising method.

## 4. Analysis of Bolt Detection Signals

Taking the high-slope anchor grouting test of Yunnan-highway as an example, the instrument is the LX-10 bolt, the sampling frequency is 10498Hz, the sampling point is 980 and the sampling interval is 4.0μs (Fig.8). The collected vibration signal is shown below in Fig 9.

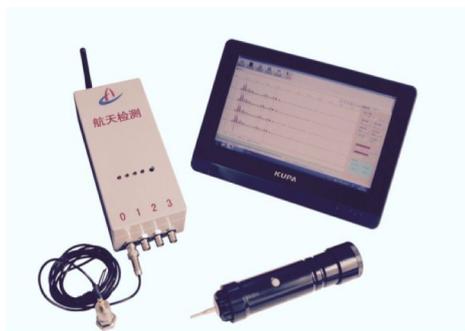
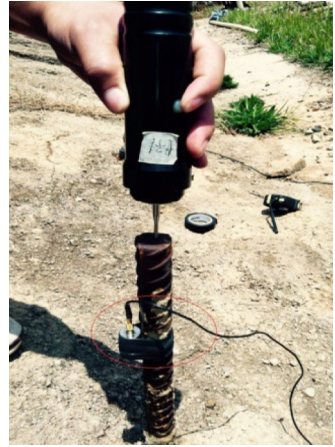

(a) Detection instrument                            (b)Testing site

**Fig.8.** Bolt detection testing instrument and site

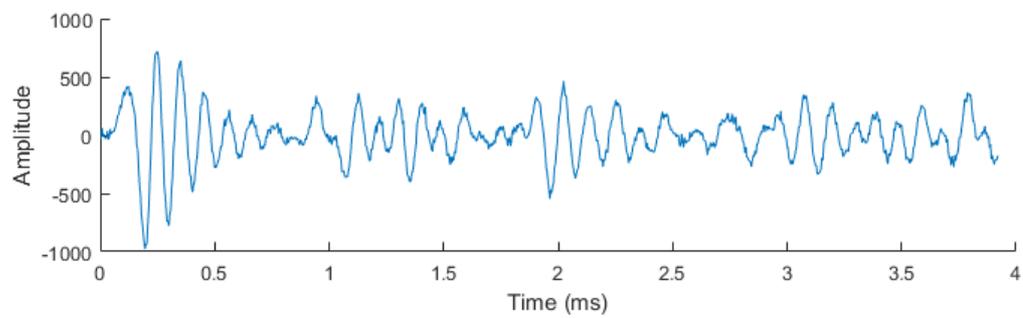

**Fig.9.** Bolt detection signal in actual engineering

The anchor detection signals in Fig.9 are decomposed by the ICEEMD method and ICEEMD-De, respectively. The results are shown in Fig. 10 and Fig. 11.

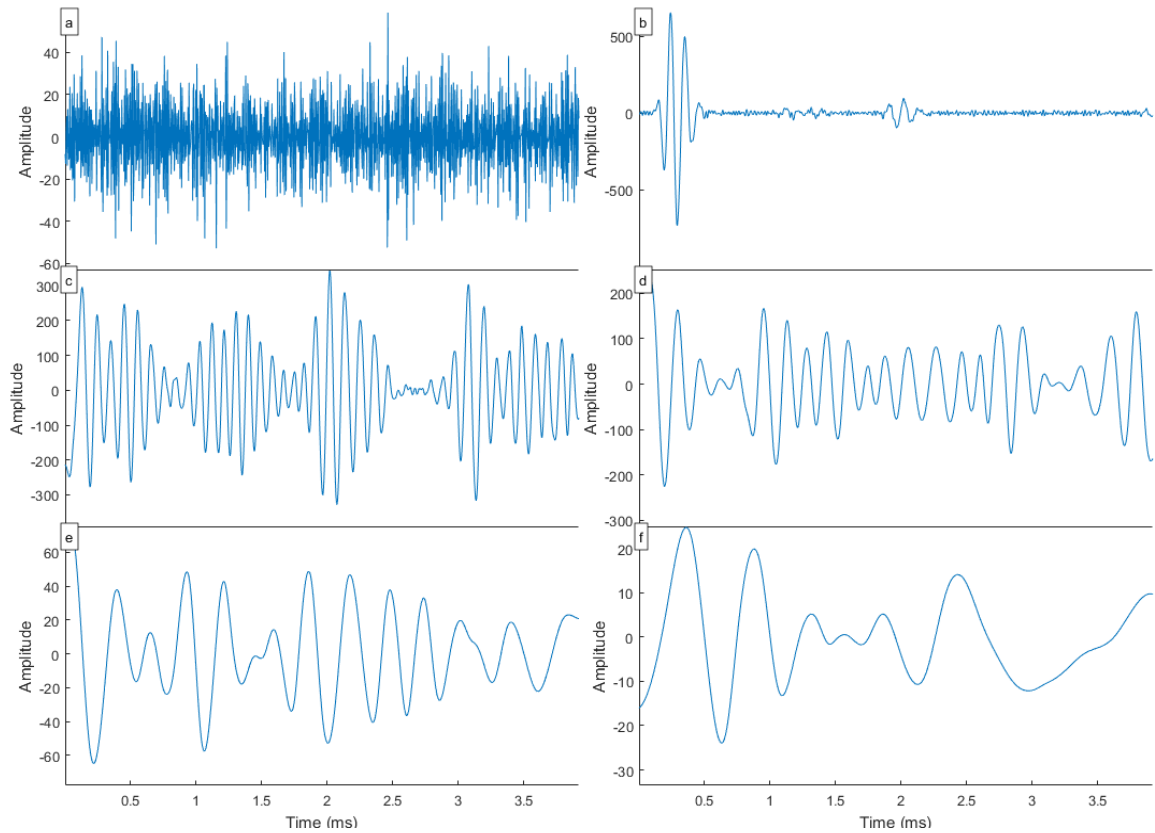

F**ig.10.**The IMF which the vibration signal after the decomposition with ICEEMD: Fig.10 (a) ~ (f) show IMF1 ~ 6 Signal decomposed with ICEEMD.

According to Fig.10, the IMF1 that is present in the high frequency noise signal, and the frequency of the IMF1 ~ 6 vibration modes gradually increase. In the modal IMF2, it is obvious that the noise signal can be seen. At 1.1ms, in the background of the bottom of the anchor reflection, the signal can be identified; however, the characteristics are not clear enough.

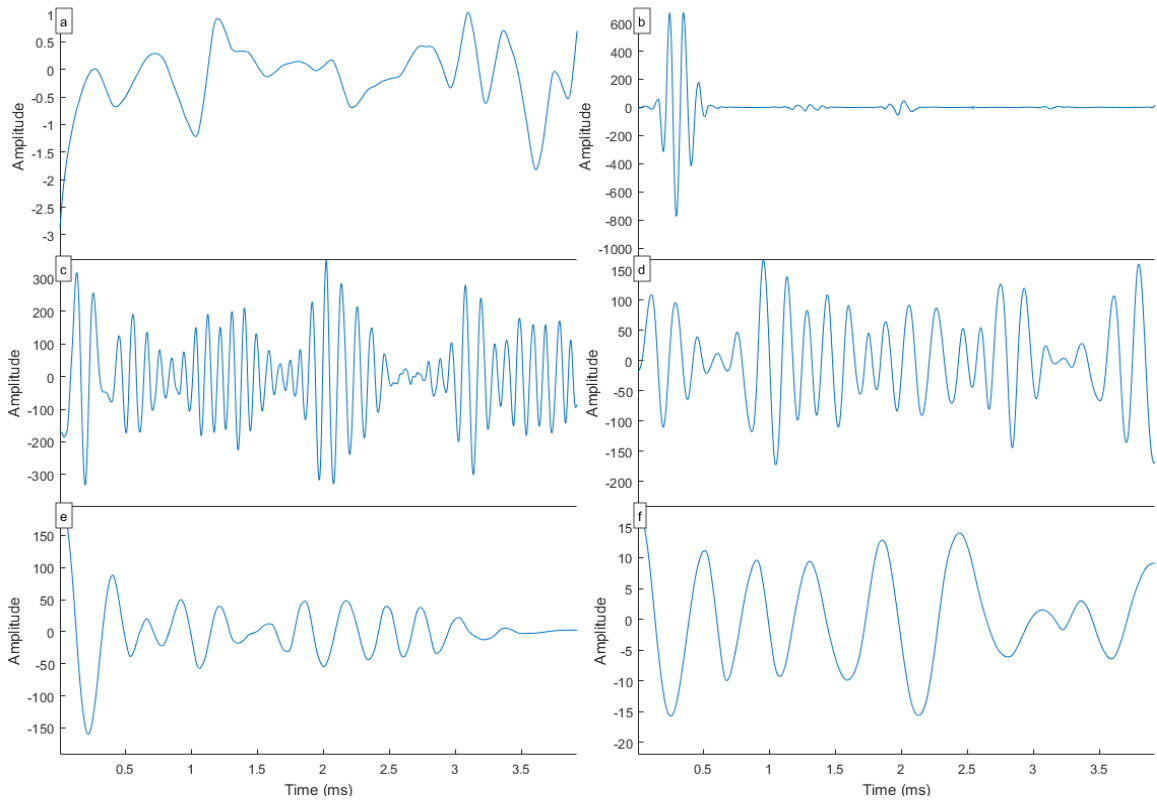

**Fig.11.** The IMF which the vibration signal after the decomposition with ICEEMD-De: (a)~(f) shows t IMF1 ~ 6 which is the signal decomposed with ICEEMD-De.

According to Fig. 11, the noise signal in each of the modes is obviously suppressed, with the frequency of the IMF1 ~ 6, the vibration modes gradually decrease, and the reflection signal of the bottom of the bolt becomes clear at 1.1ms in the modal IMF2. Fig.12 shows the initial reflection at 1.1ms in Zoom mode, and noise interference signal was significantly suppressed. As such, the de-noising effect is obvious.

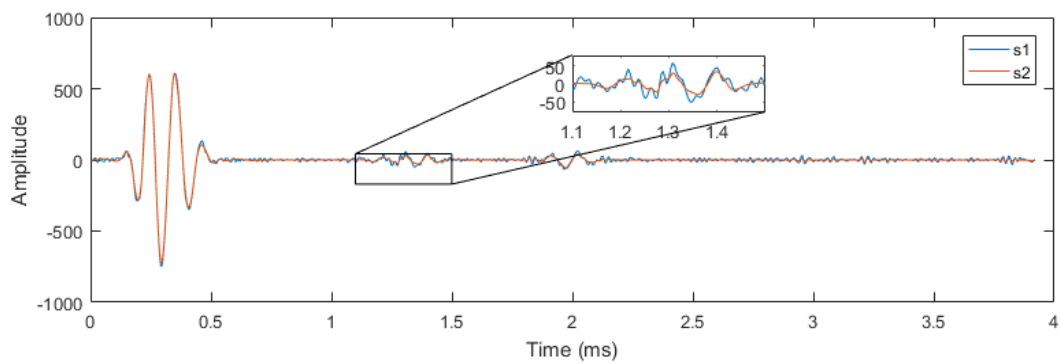

**Fig. 12.** ICEEMD IMF2vs. ICEEMD-De IMF2

## 5.Conclusions

Based on the principle of the ICEEMD decomposition, the general approximation entropy and wavelet de-noising, the ICEEMD method is introduced into the bolt detection signal analysis. The anchor detection signal is decomposed by means of using the ICEEMD, while the approximate entropy is regarded as the condition for whether or not the IMF is de-noising. Using the wavelet soft threshold de-noising technique to eliminate the noise in the IMF, the ICEEMD-De anchor signal analysis method is proposed. Based on the usage of the ICEEMD-De to analyze the vibration's analog signal and the anchor detection signal, the following conclusions have been drawn:

1) The ICEEMD method can effectively separate the vibration modal signals to IMF.

2) The ICEEMD-De method can effectively remove the interference in the vibration detection signal, and the de-noising effect is superior to that of the traditional Wavelet method.

3) The ICEEMD method is able to separate each IMF from the bolt detection signal, and the IMF can effectively identify the bolt's end reflection time.

4) In the analysis of the bolt detection signal, the ICEEMD-De is more effective at suppressing of the interference than the ICEEMD is. However, ICEEMD-De combine ICEEMD and wavelet de-noise technology. The method has many analysis steps. So the computational cost of the method is higher than traditional method.

## Acknowledgments

This research was funded by the Open Research Fund of Key Laboratory of Hydraulic and Waterway Engineering of the Ministry of Education (Grant No. SLK2017A02) and A Project Funded by the Priority Academic Program Development of Jiangsu Higher Education Institutions (Grant No.3014-SYS1401). The author wish to thank the Su Jiankun from Yunnan Aerospace Engineering Geophysical Limited by Share Ltd for providing the GPR practical detection data used in this study.